\documentclass[conference]{IEEEtran}
\IEEEoverridecommandlockouts

\usepackage{cite}
\usepackage{amsmath,amssymb,amsfonts}
\usepackage{algorithmic}
\usepackage{graphicx}
\usepackage{textcomp}
\usepackage{xcolor}
\def\BibTeX{{\rm B\kern-.05em{\sc i\kern-.025em b}\kern-.08em
    T\kern-.1667em\lower.7ex\hbox{E}\kern-.125emX}}

\usepackage{tabularx}
\usepackage[colorlinks,allcolors=blue]{hyperref}

\usepackage{comment}

\usepackage{multirow}
\usepackage{xspace}
\usepackage{todonotes}
\usepackage{booktabs}
\usepackage{mathrsfs}

\usepackage{url}
\usepackage{subcaption}

\usepackage[inkscapelatex=false]{svg}

\newcommand\acronym{\textsc{EVolve}\xspace}

\makeatletter 
\newcommand{\linebreakand}{%
  \end{@IEEEauthor0halign}
  \hfill\mbox{}\par
  \mbox{}\hfill\begin{@IEEEauthorhalign}
}
\makeatother 

\begin{document}
\title{\acronym: a Value-Added Services Platform for Electric Vehicle Charging Stations}

\author{
\IEEEauthorblockN{
Erick Silva\IEEEauthorrefmark{2},
Tadeu Freitas\IEEEauthorrefmark{1},
\thanks{Tadeu Freitas was supported by the grant 2021.04529.BD (FCT).}
Rehana Yasmin\IEEEauthorrefmark{3}, 
Ali Shoker\IEEEauthorrefmark{4},\\
Paulo Esteves-Verissimo\IEEEauthorrefmark{5}}
\IEEEauthorblockA{
King Abdullah University of Science and Technology (KAUST) and\\Faculty of Science of University of Porto (FCUP),\\
Email:\{erick.silva\IEEEauthorrefmark{2}, 
rehana.yasmin\IEEEauthorrefmark{3}, 
ali.shoker\IEEEauthorrefmark{4}, 
paulo.verissimo\IEEEauthorrefmark{5}\}@kaust.edu.sa,\\
Email:{tadeufreitas@fc.up.pt\IEEEauthorrefmark{1}
}
}}
\maketitle

\begin{abstract}
A notable challenge in Electric Vehicle (EV) charging is the time required to fully charge the battery, which can range from 15 minutes to 2-3 hours. However, this idle period for the EV presents an opportunity to offer time-consuming or data-intensive services such as vehicular software updates. ISO 15118 referred to the concept of Value-Added Services (VASs) in the charging scenario, but it remained underexplored in the literature. Our paper addresses this gap by proposing \acronym, the first EV charger compute architecture that supports secure on-charger universal applications with upstream and downstream communication. The architecture covers the end-to-end hardware/software stack, including standard API for vehicles and IT infrastructure.  We demonstrate the feasibility and advantages of \acronym by employing and evaluating three suggested value-added services: vehicular software updates, security information and event management (SIEM), and secure payments. 
The results demonstrate significant reductions in bandwidth utilization and latency, as well as high throughput, which supports this novel concept and suggests a promising business model for Electric Vehicle charging station operation.

\end{abstract}


\section{Introduction}
\label{sec:intro}
With the increasing adoption of EVs, the demand for advanced, secure, and scalable charging infrastructure has become essential \cite{iea_global_ev_outlook_2024}\cite{avidthink_aws_automotive_2020}. Governments and industries worldwide are heavily investing in EV technologies to reduce carbon footprints and transition towards sustainable mobility solutions \cite{CARB_ZEV} \cite{doe_what_electrification}. However, EVs' prominence brings its own challenges, particularly related to cybersecurity risks in both vehicles and charging systems, limited and inconsistent charging infrastructure,  and frustrating user experience.



Standards such as ISO 15118 \cite{iso15118-1-2019} \cite{iso15118-2-2014} \cite{iso15118-20-2022} have been introduced to address these security and communication challenges. Still, significant gaps remain, especially in supporting data-rich and secure interactions between vehicles, chargers, and cloud-based services \cite{ChargePoint_VDV261} \cite{avidthink_aws_automotive_2020}. Additionally, range anxiety and idle charging time remain major issues for EVs, impacting adoption rates. 
Drivers often worry about reaching a charging station before running out of charge, only to encounter long waits due to occupied or overstayed chargers. This mix of uncertainty and delay makes the overall charging experience frustrating and unreliable. However, enabling productive or engaging activities during this period could shift user perception, reduce concerns, and benefit the industry by enhancing the overall charging experience.

Traditional EV chargers primarily focus on energy transfer while neglecting their potential to support advanced, data-centric applications. Cellular communications, often used in vehicles, are unreliable and add extra costs, making them less suited for data-heavy applications in charging scenarios \cite{broadlinc_cellular_data_drawbacks}. This underscores the need for a solution that addresses security and operational inefficiencies while enhancing user experience.

This work introduces \acronym, a platform designed to elevate EV chargers from energy dispensers to robust Vehicular Edge Computing (VEC) devices. \acronym brings a novel concept transforming the EV charging ecosystem by supporting Value Added Services (VASs) in the charging environment. By enhancing data communication and security and introducing innovative services, \acronym addresses critical gaps in the existing infrastructure.


Our contributions are as follows: 1) the \acronym, a novel VEC architecture that transforms EV chargers into secure, data-driven hubs bridging OT and IT, compliant with ISO 15118; 2) the integration of VASs such as software updates, SIEM, and secure payments; 3) the practical implementation and feasibility demonstration of \acronym; and 4) empirical performance evaluation showing significantly lower latency and higher throughput compared to 4G/5G, with up to a seven-fold improvement over 5G in high-frequency tasks like secure payments, enabled by wired Power Line Communication (PLC) as specified by ISO 15118. Moreover, our architecture is the base for our work on Security Orchestration, Automation and Response for EVs on \cite{freitas2024evsoar}, showcasing the platform's flexibility and practicality.

The remainder of this paper is organized as follows: Section 2 reviews the related work that inspired and informed our research. In Section 3, we explore the challenges and opportunities in the domain and provide a detailed overview of our proposed architecture. Section 4 focuses on the implementation and evaluation of three VASs, detailing how the identified challenges were addressed. This section also includes an analysis of the evaluation results and considerations of system overhead and requirements. Finally, Section 5 concludes the paper, summarizing key findings.

\section{Related Works}
\label{sec:related}


The ISO 15118 (first published in 2014) serves as the foundational standard for enabling high-level communication between EVs and Supply Equipment Communication Controller (SECC). It also briefly highlights the importance of having VASs in the future, that leverage the edge-like data and compute capabilities of EV charger. Nevertheless, this topic has not been addressed deeply in literature, and no generic architectures have been proposed, unless primitive use-cases like online payments.

VEC brings computational and storage resources closer to vehicles, addressing limitations in cloud-dependent systems by reducing latency and enhancing real-time processing. As highlighted in \cite{meneguette2021vehicular} and \cite{caliandro2020multi}. Edge computing enables a wide range of applications, including software updates \cite{shoker2023scalota}, map download, and predictive maintenance \cite{khorsravinia2017integrated}, all of which are essential for modern EV infrastructure.

Previous studies \cite{rong2024joint} \cite{zhong2024joint} have explored the role of EVs as both energy prosumers and computing nodes, optimizing charging, computation, and travel costs through coordinated management of EVs, the Power Distribution Network, and the Computing Power Network, a fusion of EV and edge server. While these works primarily focus on optimizing EV-side operations, our research takes a different approach by emphasizing the SECC and edge servers, ensuring that the underlying infrastructure can support such capabilities. Furthermore, we adopt the concept of VASs, facilitating the deployment of these technologies rather than solely optimizing their operation. Consequently, our work complements rather than overlaps with these prior studies.


Since its introduction in 2013 \cite{rodriguez2013added}, VAS has evolved to expand the value proposition of EV charging stations. Software updates represent one of the earliest practical applications, addressing the underutilization of EV chargers as service delivery points. By leveraging the proximity of chargers to vehicles, works such as ScalOTA \cite{shoker2023scalota} have developed secure, end-to-end update protocols, utilizing idle charging time to ensure efficient and secure firmware updates for Electronic Control Units (ECUs). This approach reduces reliance on workshops and enhances convenience for both Original Equipment Manufacturers (OEMs) and users, highlighting the role of VEC in modernizing the EV ecosystem. Furthermore, the industry sees edge computing as an essential component for the delivery of comfort and convenience applications \cite{avidthink_aws_automotive_2020}, emphasizing the importance of integrating VASs into the EV infrastructure. \acronym generalizes the OTA update application of ScalOTA to cover more applications and support OEM-agnostic vehicles.

The integration of VASs into the EV charging ecosystem has been driven by efforts to utilize idle charging time more effectively. One prominent example is the VDV resolution 261 \cite{CarMediaLab_VDV261}, which introduced a VAS to precondition electric buses during charging in Germany. This service, compliant with ISO 15118, ensures that bus interiors are climatized, enhancing passenger comfort and operational efficiency while conserving battery power for driving. 

While ISO 15118 provides a solid foundation for secure EV communication, its gaps in VAS practical implementation and scalability highlight the need for complementary solutions. Edge computing addresses these limitations by providing a distributed data processing architecture that supports diverse applications with enhanced security and efficiency. By leveraging edge-based platforms like Edgex \cite{edgexDocumentation} and Baetyl \cite{Baetyl2024}, which integrate microservices and advanced communication protocols, it is possible to deploy secure, scalable, and user-centric VASs in the EV ecosystem. \acronym's implementation benefits greatly from the integration of these platforms to facilitate building a solid end-to-end solution.

\section{\acronym Architecture}
\label{sec:arch}
\subsection{Overview}
\acronym is a VEC-based platform designed to bridge the gap left by ISO 15118 in defining and supporting VASs within the EV charging ecosystem. By leveraging EV chargers' communication and computation capabilities, the platform enables VASs to create a secure, robust environment for running diverse services. \acronym also follows modular architecture, standard APIs, and protocols to ease its integration and facilitate building new applications. \acronym emphasizes both security and performance, addressing the lack of a systematic approach to functional, performance, and security requirements for VASs in the literature. 

\subsection{Challenges and Opportunities}
Although there are huge efforts to reduce the charging time and optimize battery utilization, extending the EV chargers with data, computing, and communication capabilities entails huge opportunities. First, it reduces the psychological feeling of ``time waste''. Second, it improves the utilization of cellular infrastructure resources (bandwidth and Mobile Edge Compute (MEC)), interference, and reduces the latency of services. Third, automotive applications can now benefit from the diverse data sharing of several fleets and models at the EV charging station. Fourth, it enables a wide set of new services in the \textit{connected} automotive ecosystem, which was impossible with the constrained vehicle capabilities or overwhelming the IT infrastructure. This opens new business opportunities where EV charging operators can support automotive applications through offloading heavyweight computing or communication to the EV chargers or downloading huge amounts of data (e.g., maps, software updates, etc.).

Unfortunately, providing the above opportunities raises operational and security challenges. The operational ones are mainly related to the complex and rich modern automotive ecosystem spanning vehicles, EV chargers, Infrastructure, OEMs, etc. This requires an architecture that has wide support and easy integration. The security challenges are obvious because EV chargers in the open air are typically less protected than classical IT infrastructure and expand the attack surface. In addition, sharing data can have privacy concerns. This requires a platform that supports state-of-the-art security capabilities, mitigations, and data privacy techniques. 

The above considerations inspired us to design \acronym with an emphasis on the trade-offs between operation, performance, and security. We achieved this by making the following design decisions: (1) adhering to standard interfacing and communication protocols to support different suppliers across the ecosystem; (2) adopting a layered architecture that enables the integration of new hardware components (e.g., GPUs, FPGAs, DPUs, TEEs, etc.), software modules, and communication methods to support extensibility without impacting application layer development; and (3) implementing best practices for communication and ensuring support for computing security.

\subsection{Architecture}

\begin{figure}[htp]
    \centering
    \includegraphics[width=\linewidth]{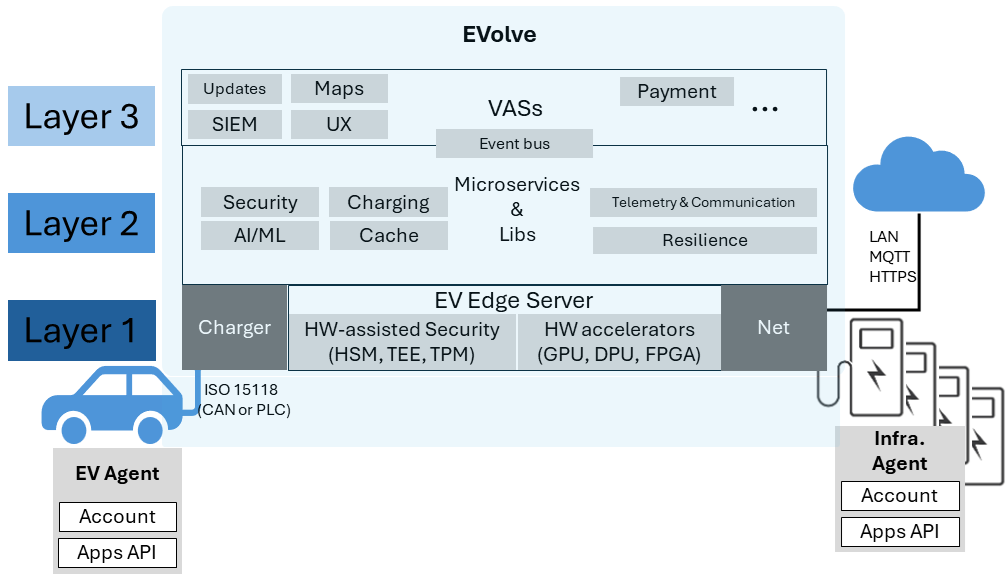}
    \caption{Architecture of \acronym.}
    \label{fig: EV charger Architecture}
\end{figure}

\acronym's architecture integrates edge computing capabilities into the SECC, enabling it to function as a VEC device. Figure \ref{fig: EV charger Architecture} illustrates this architecture, organized into three distinct software layers that collectively address the platform's usability and cybersecurity requirements. These layers support enriching and processing data locally or securely forwarding it to reduce user data consumption by leveraging the existing charger link.

\subsubsection{Layer 1 – Edge Server}
The first layer includes the low-level firmware for the charger, hardware accelerators, and the EV Edge Server. This layer provides universal and flexible \textit{data handling} by supporting diverse data formats and enabling seamless communication between vehicle-side and cloud infrastructures. The edge server manages data reception from the vehicle and provides interfaces to resources such as hardware-assisted security and accelerators, ensuring integration and functionality for VASs.

\subsubsection{Layer 2 – Microservices and Libraries}
The second layer incorporates critical \textit{microservices and libraries} designed to meet stringent cybersecurity requirements and optimize the performance of VASs. This layer encompasses essential modules such as \textbf{Security}, which includes cryptographic libraries, protocols, and secure hardware module drivers for secure operations, storage, and sensitive data computation; \textbf{Charging Stack}, ensuring reliable charging operations with high availability; \textbf{AI and Machine Learning}, featuring AI processing software libraries to support offloaded data analytics and inference, enhancing VAS efficiency; \textbf{Cache Storage}, providing efficient caching to minimize latency and improve availability; \textbf{Telemetry \& Communication}, collecting data and ensuring secure cloud export; and \textbf{Resilience}, delivering fault tolerance and recovery through monitoring, redundancy, and proactive failure handling to guarantee system reliability.

\subsubsection{Event Bus – The Link Between Layers 2 and 3}
The event bus facilitates communication between microservices and VASs, ensuring transparency and efficiency. This setup is critical for implementing \textit{controlled access} to VAS, as access control lists restrict resource access based on application roles and criticality levels. This mechanism protects sensitive modules while preventing performance degradation, ensuring secure and efficient operations.

\subsubsection{Layer 3 – Value-Added Services (VASs)}

The top layer delivers advanced functionalities, including \textbf{Vehicular Software Updates} for over-the-air firmware/software updates, \textbf{SIEM} integrating telemetry and analytics for threat monitoring and mitigation, and \textbf{Secure Payments} supporting charging and VAS transactions for untrusted stakeholders, alongside maps, entertainment, and user experience enhancements, leveraging the lower layer's capabilities to meet usability and cybersecurity requirements while ensuring scalability and adaptability.

\subsubsection{Application Programming Interface (API)}

We now explore the EV charger interaction capabilities with external entities, specifically vehicles and cloud systems. Ensuring accessibility and user-friendliness for vehicle-side and cloud-side developers is critical to this work. The proposed API bridges these connections while prioritizing security, efficiency, and ease of integration.

\paragraph{Vehicle-Side Communication}
 The communication with the charger begins with the SECC Discovery Protocol (SDP), a UDP-based mechanism that identifies chargers on the same network. The charger responds to the SDP with its IP address and port number, enabling the vehicle to initiate secure communication. This interaction marks the activation of the vehicle-side API, which handles the Service Negotiation Protocol (SNP). During the SNP, the vehicle and charger exchange parameters and values for the services offered.
The communication platform is built on encrypted data exchange, adhering to TLS standards with EDH for key exchange, as specified by ISO 15118. This ensures secure communication, whether conducted over wired connections, such as charging cables, or wirelessly.
To simplify and optimize vehicle-to-charger communication, we propose an API at the AUTOSAR architecture application layer \cite{furst2009autosar}. The API is activated when the charging service is initiated following the successful establishment of a TLS connection between the vehicle and charger.
Once activated, the API engages with the SNP to process the services available on the charger (SECC) and any parameters required for their operation. Designed to be lightweight, the API ensures minimal impact on the vehicle’s primary functions, maintaining its efficiency. 

\paragraph{Cloud-Side Communication}


Seamless, encrypted communication between the EV charger and the cloud, utilizing HTTPS and MQTTS, is crucial for VASs requiring cloud-hosted components or heavier processing, supporting: offloading computationally intensive tasks, synchronizing logs and data analytics for advanced VASs like SIEM, predictive maintenance and grid load balancing, and secure data export for third-party applications. This cohesive vehicle-side and cloud-side API integration allows the platform to act as a robust intermediary, bridging edge and centralized systems while upholding security, performance, and scalability.

\subsubsection{Integration with ISO 15118 and Beyond}
While ISO 15118 primarily focuses on vehicle-to-charger and grid communication, \acronym extends its scope by integrating services requiring data exchange among VASs, such as SIEM-triggered software updates. These interactions demand heightened security and trust management, which are addressed through controlled access and prioritization of critical services. Access control lists implemented within the event bus maintain service integrity and availability, ensuring the uninterrupted operation of essential functionalities.

\section{Evaluation}
\label{sec:eval}

\subsection{Evaluation Methodology}
The platform evaluation consists of two primary components: an analytical assessment of the platform’s requirements to verify compliance with our proposed requirements and an empirical evaluation of memory (RAM) usage and footprint size to assess the platform’s impact on hardware resources. This dual approach provides a comprehensive understanding of the platform's adherence to defined requirements and resource efficiency.

From the perspective of added services, we empirically evaluate three distinct classes of applications: software updates, SIEM, and secure payments. Each application class presents unique data characteristics and communication requirements within the platform. Vehicular software updates involve large downstream data transfers, representing substantial data downloads. The SIEM service, as a security solution, alternates between (1) pushing large log files and (2) pulling small Federated Learning (FL) parameters; FL is used as a privacy-preserving technique for learning. Finally, the secure payments VAS processes small, frequent data transactions, requiring high-frequency communication between the vehicle and edge to demonstrate the platform’s reduced latency capabilities. Together, these applications provide a comprehensive view of the platform’s performance across diverse communication demands. The following sections will explain the scenario created to experiment with our services. Then, we will go through each of the evaluated VASs, analyze the overhead in the system, and finally go through the solutions used to fulfill the requirements of our architecture.

\subsection{Experimental Setting}



Our environment consisted of an emulation made on Emulab \cite{emulab}. In this emulation, machines were configured to represent variations of systems with PLC and cellular. These configurations allow us to evaluate three communication cases via PLC: EVolve10, EVolve100, EVolve1G, and two cases using cellular networks: 4G and 5G, as summarized in Table \ref{tab:network_performance}. PLC is used for Vehicle-to-Charger communication, while LTE represents communication between the vehicle or EV charger and the cloud. 
The selected PLC speeds range from 10 Mbps, which aligns with the current HomePlug Green PHY standard \cite{homeplug_green_phy}, to 100 Mbps, which reflects ongoing industry advancements, and finally to 1 Gbps, the target of future vendor developments. These speed variations provide insights into how different communication rates impact system performance.

To measure round-trip time (RTT) latency, we recorded a local timestamp, sent a message to a server, and calculated the time difference upon receiving the response. For each payload size and media type, 300 measurements were recorded to ensure consistency and reliability. Payload sizes ranged from 1 KB to 1 GB for the vehicle (client), with responses from the charger (server) varying from simple acknowledgments (ACK) to 1 GB. The connection metrics, including Throughput (TPS), Latency, and Packet Loss Rate (PLR), for the physical layer (PHY) using PLC were derived from ScalOTA \cite{shoker2023scalota} and the HomePlug Green Phy standard. Additional network parameters for each connection type were referenced from recent empirical studies \cite{digitaltrends_5g_speed} \cite{ding2021understanding} to ensure the evaluation accurately reflects real-world conditions and network performance. The PLR, critical for assessing connection stability, is summarized together with the other parameters in Table \ref{tab:network_performance}.

\begin{table}[h!]
    \centering
        \caption{Network Performance Comparison for Different Connections}
    \begin{tabular}{|c|c|c|c|c|}
        \hline
        \textbf{Configuration} &\textbf{Connection} & \textbf{TPS (Mbps)} & \textbf{Latency (ms)} & \textbf{PLR} \\
        \hline
        EVolve10  & PHY (PLC) & 10 & 2 & 0 \\
        EVolve100 & PHY (PLC) & 100 & 2 & 0 \\
        EVolve1G  & PHY (PLC) & 1000 & 2 & 0 \\
        4G & Wireless (4G) & 30 & 36 & 0.2 \\
        5G & Wireless (5G) & 100 & 17 & 0.2 \\
        \hline
    \end{tabular}

    \label{tab:network_performance}
\end{table}

\begin{figure}[htp]
    \centering
    \includegraphics[width=0.9\linewidth]{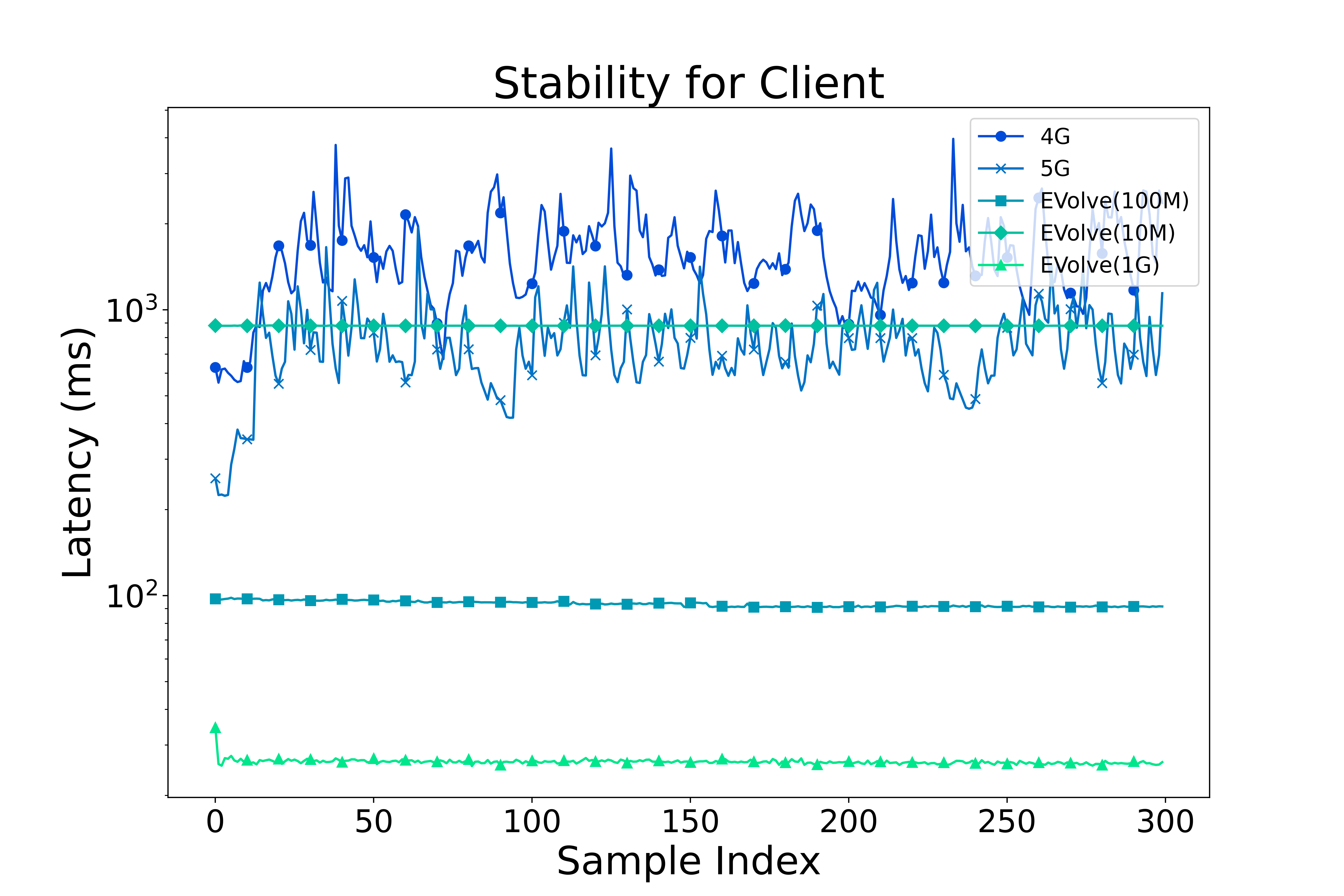}
    \caption{Stability for Client.}
    \label{fig:stability-client}
\end{figure}

Following the measurements, the collected points were plotted in a time series manner for visualization. This first analysis, shown in Fig. \ref{fig:stability-client}, illustrates the higher stability of wired connections (EVolve10, EVolve100, EVolve1G).

\subsection{Value-Added Services}

\subsubsection{Vehicular Software Updates}
\begin{figure}[htp]
    \centering
    \includegraphics[width=0.8\linewidth]{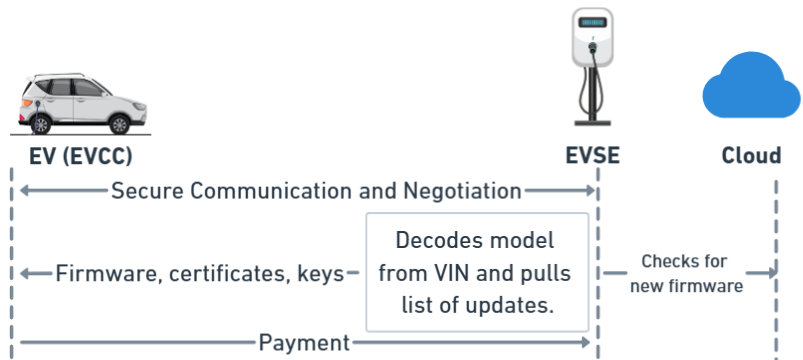}
    \caption{Vehicular Software Update Service Flow.}
    \label{fig:update-chain}
\end{figure}

Fig. \ref{fig:update-chain} shows the flow for the software updates VAS. This service uses a pub/sub mechanism to receive notifications of available firmware updates from an Image Repository. The EV charger then downloads updates securely from the cloud if they are not cached, authenticates them, and implements version control and rollback mechanisms. This architecture minimizes network reliance and ensures that updates are securely delivered and applied without compromising the system.

\begin{figure}[htp]
    \centering
    \includegraphics[width=0.8\linewidth]{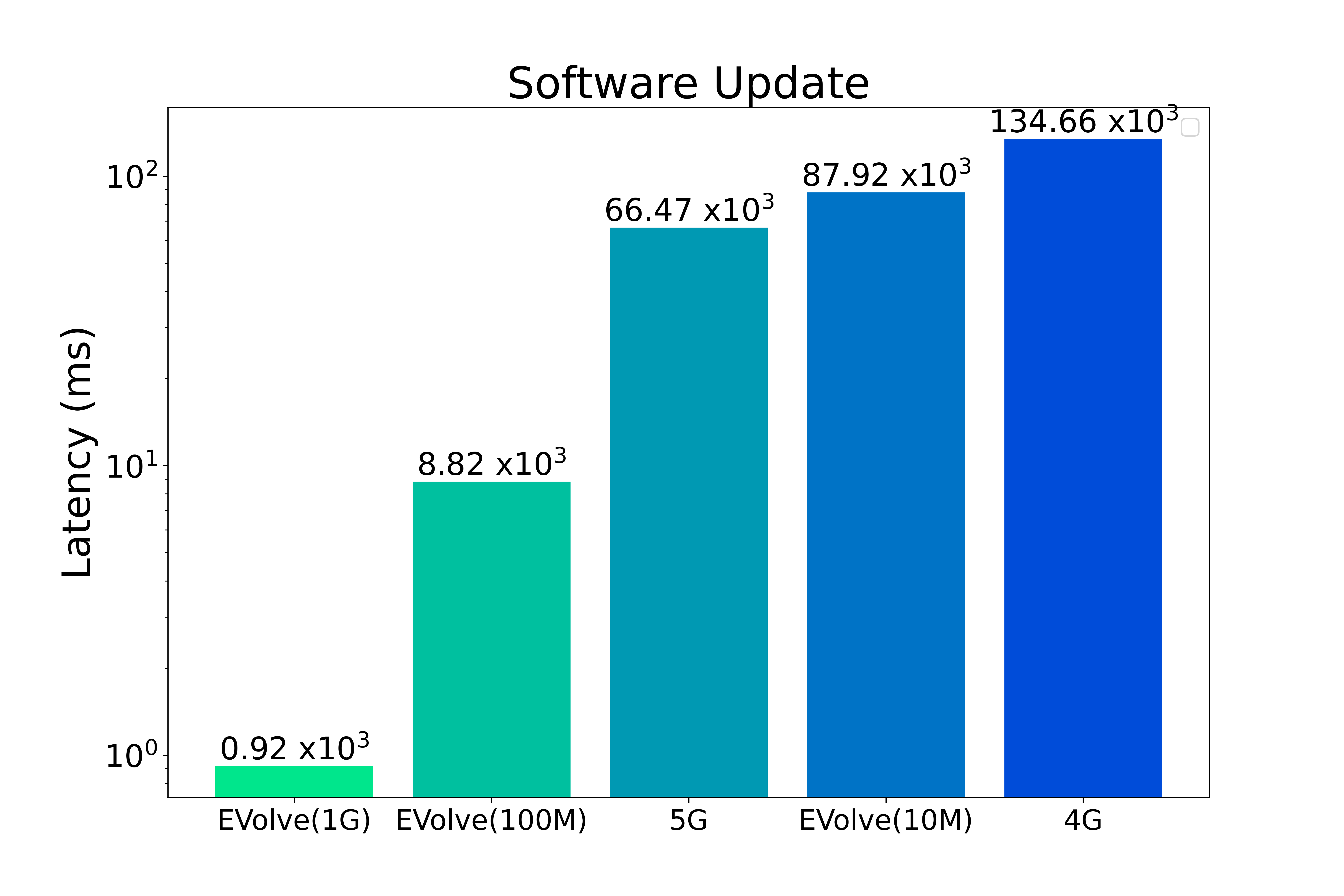}
    \caption{Latency for Software Updates.}
    \label{fig:eva-updates}
\end{figure}

In this scenario, we evaluated system performance by emulating an initial request with a 1 KB payload, followed by a substantial 100 MB software update download, representing typical downstream communication. This 100 MB payload aligns with the dimensions used in ScalOTA. However, unlike ScalOTA, the test did not incorporate cache storage, allowing for a direct comparison of raw download performance.

Figure \ref{fig:eva-updates} shows that the current PLC achieves lower latency than 4G but does not surpass 5G performance. However, the next-generation 100 Mbps PLC demonstrates significantly lower latency, making it a highly competitive choice, particularly for applications requiring large, uninterrupted data downloads, such as vehicular software updates. These findings highlight the advanced capabilities of new PLC technology for high-speed data transfer. The even more advanced 1 Gbps PLC technology offers exceptional performance for this case, being almost 10 times faster than the 100 Mbps one.

\subsubsection{SIEM}
\begin{figure}[htp]
    \centering
    \includegraphics[width=0.8\linewidth]{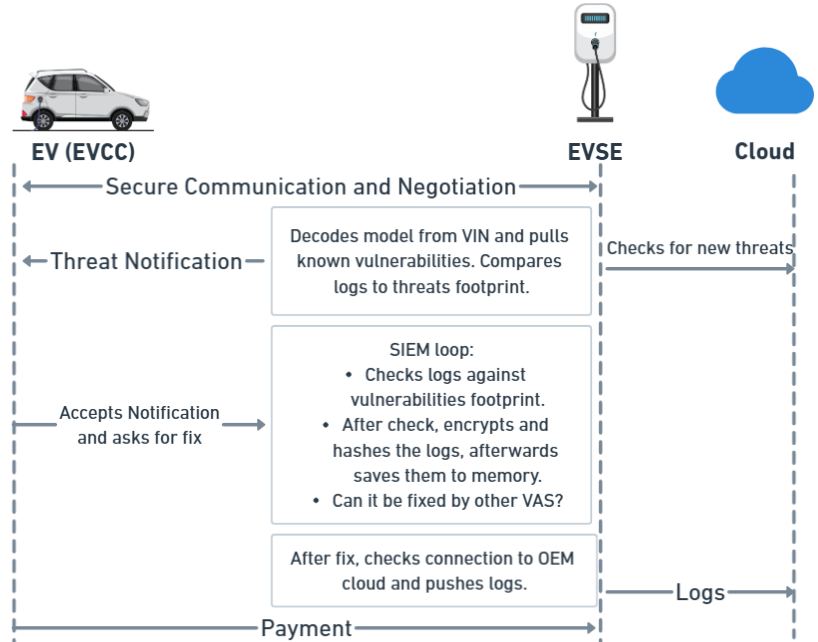}
    \caption{SIEM Service Flow.}
    \label{fig:siem-chain}
\end{figure}

The SIEM process, depicted in Fig. \ref{fig:siem-chain}, relies on an external agent for off-board analysis of logs transmitted securely from the vehicle via the EV charger. This design avoids the risks of running real-time Intrusion Prevention Systems (IPS) within vehicles, where false positives could disrupt operation. The SIEM service on the charger analyzes the logs using correlation rules, machine learning models for anomaly detection, and endpoint protection mechanisms to detect vulnerabilities or threats. This architecture enhances bandwidth efficiency and robustly responds to potential cybersecurity risks.

\begin{figure}[htp]
    \centering
    \includegraphics[width=0.8\linewidth]{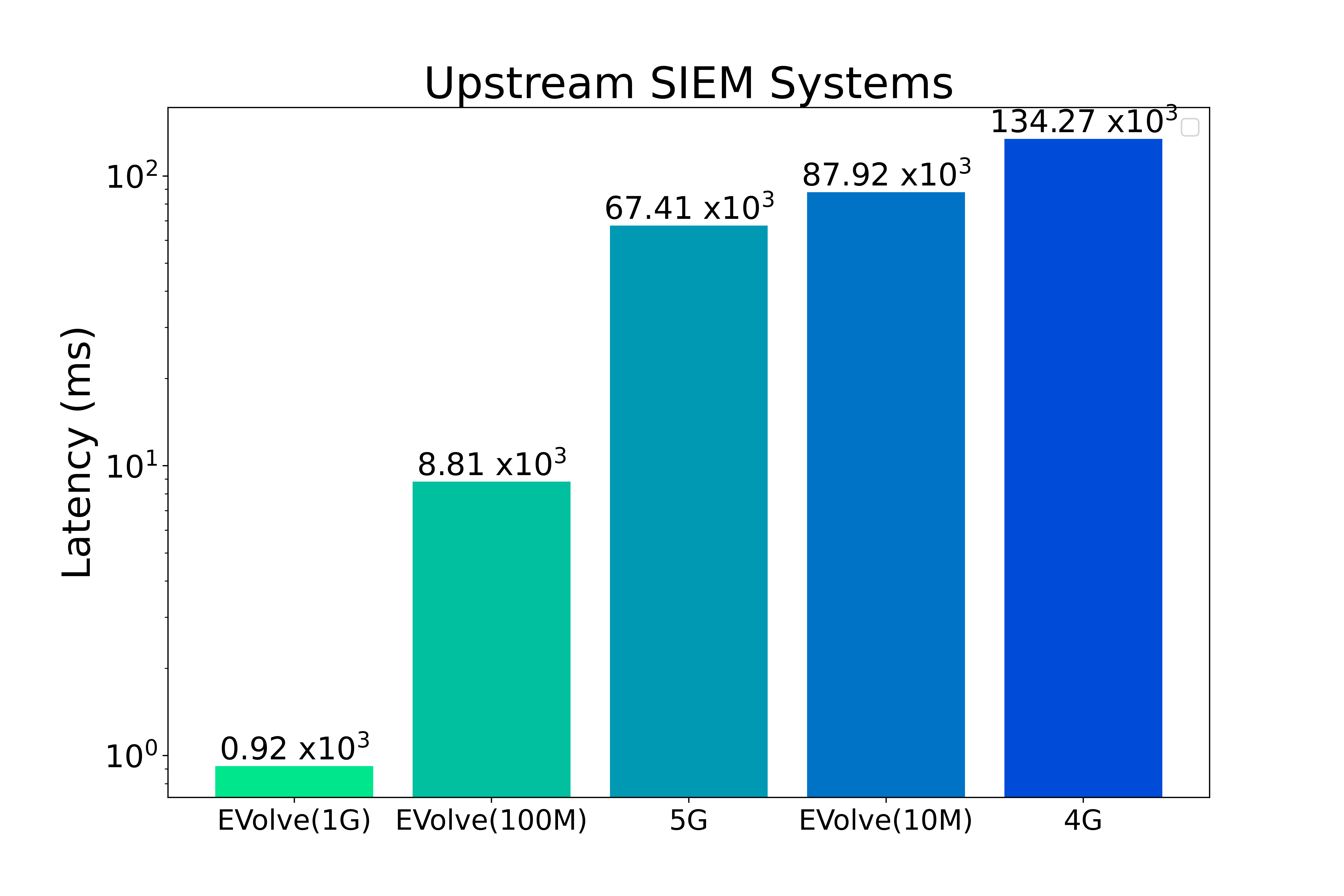}
    \caption{Latency of SIEM Operations.}
    \label{fig:eva-siem}
\end{figure}

For the SIEM VAS, which manages both large data logs and smaller FL parameters, we emulated communication based on a CAN dataset, representing 10 minutes of data totaling approximately 85 MB \cite{b_can_intrusion_dataset}. In this test, the client sent a 100 MB log file and received a simple ACK from the server, emulating the log upload to the SIEM service. We also tested a small download of 1 MB to evaluate downstream performance in receiving FL parameter updates. However, this case had the same result as the software update. Thus, we chose to omit it.

\paragraph{Uploading Logs to Server}

Figure \ref{fig:eva-siem} illustrates the latency results for upstream communication in the SIEM VAS; this is the case when we push logs to the OEM. Notably, the 5G connection outperformed the current PLC generation. However, it still did not match the exceptionally low latency and high stability of the next-generation 100 Mbps PLC, which reduced latency by nearly 90\% compared to 5G. The even more advanced 1 Gbps PLC standard shows outstanding improvements in both speed and stability, making it highly effective for efficiently transmitting large logs with minimal delay.



\subsubsection{Secure Payment}
\begin{figure}[htp]
    \centering
    \includegraphics[width=0.8\linewidth]{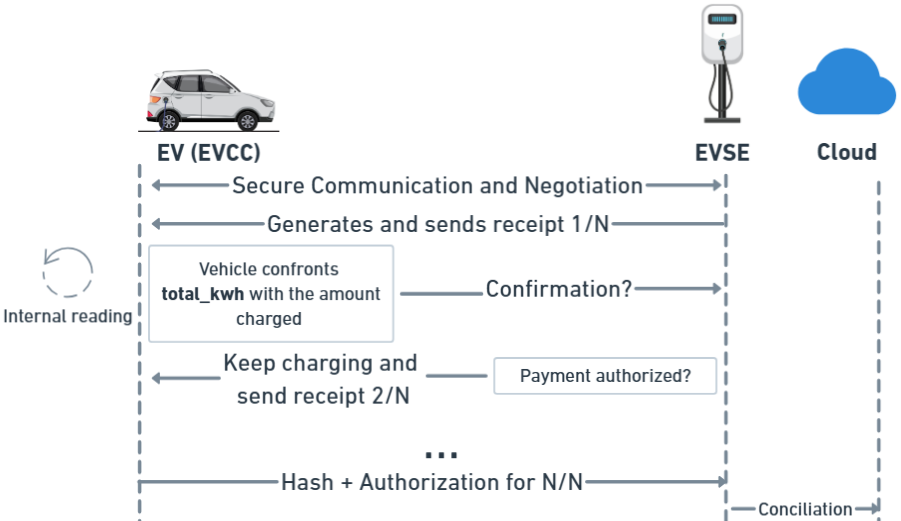}
    \caption{Secure Payment Service Flow.}
    \label{fig:payment-chain}
\end{figure}

Fig. \ref{fig:payment-chain} illustrates the secure payment process, where the EV charger facilitates real-time reconciliation with financial institutions. The charger handles frequent, small-scale transactions, ensuring confidentiality and integrity through encrypted communication. This approach provides a secure and efficient alternative to traditional payment methods, enhancing the user experience during charging sessions.

The Secure Payment VAS addresses security concerns by facilitating sensitive payment transactions between two untrusted parties, with security as its primary goal rather than performance. Nevertheless, in this evaluation, we compare the latency of a naive payment scenario to that of micropayments, acknowledging that the enhanced security offered by the Secure Payment VAS is a critical benefit beyond mere performance metrics.

\paragraph{Case 1: Naive Payment}

In the first scenario, we emulate a naive payment process where payment is completed either before or after the charging session. This evaluation focuses solely on the reconciliation phase, represented by the payload sent at the end of the Secure Payment protocol. 

\begin{figure}[htp]
    \centering
    \includegraphics[width=0.8\linewidth]{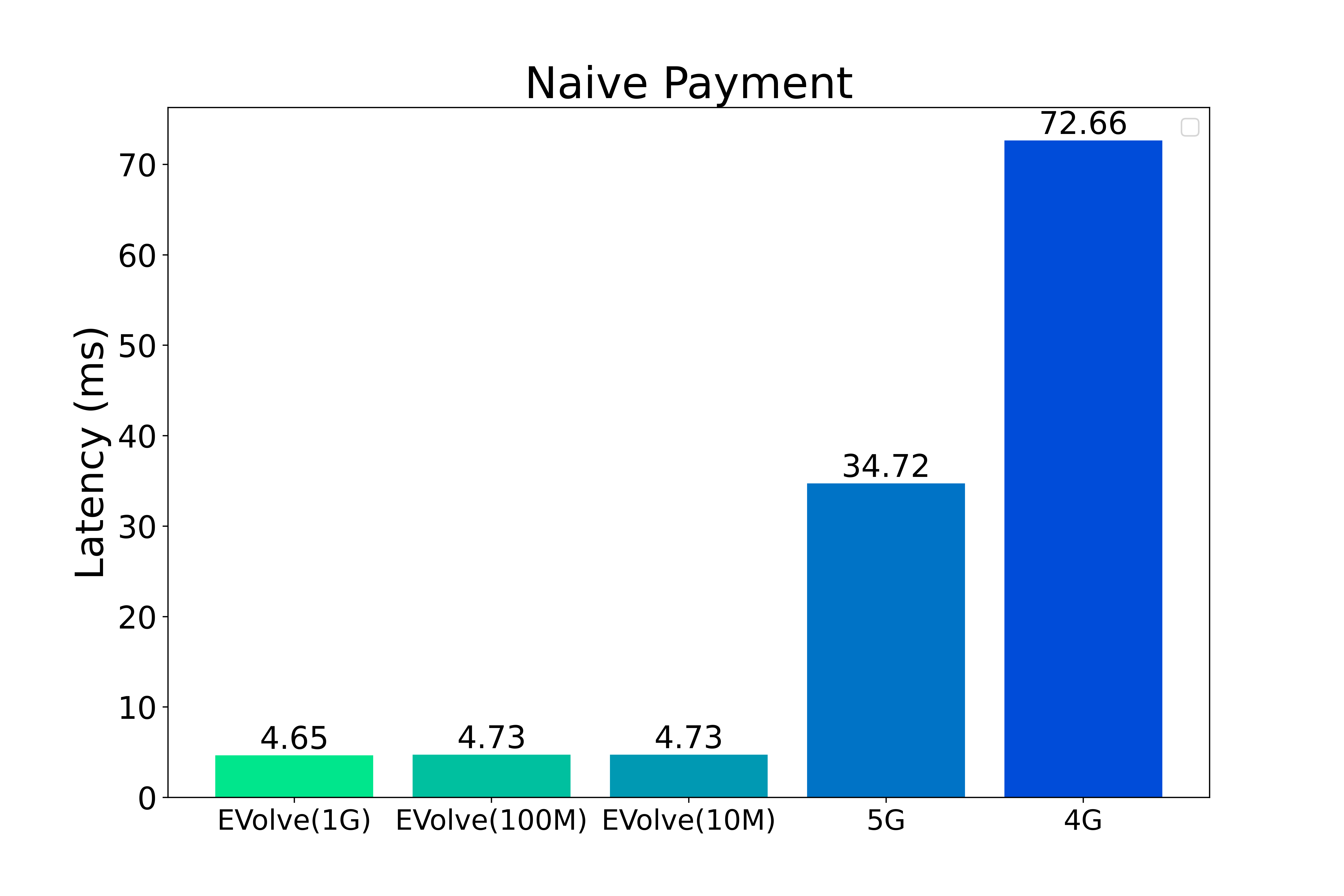}
    \caption{Latency for Simple one-time Payment.}
    \label{fig:eva-naive-payment}
\end{figure}

While the actual payload size for reconciliation is 268 bytes, due to measurement limitations, we rounded this up to 1 KB in the latency plots shown in Fig. \ref{fig:eva-naive-payment}.

The results reveal that for small payloads, all PLC connections significantly outperform 5G, achieving approximately seven times better latency. This indicates PLC’s superior efficiency in low-payload transactions, where minimal delays are essential.

\paragraph{Case 2: Micropayments}

In this scenario, we emulate the Secure Payments process, where the charging session is divided into multiple small transactions, or "bursts," each with its own micro-receipt. Two types of payloads are used here: the micro-receipt sent by the charger and the payment authorization sent by the vehicle.
The empirical sizes of these payloads are 97 bytes for the charger’s micro-receipt and 119 bytes for the vehicle’s confirmation. For consistency in measurement, both were rounded up to 1 KB, with each communication consisting of a 1 KB request and a 1 KB reply.



\begin{figure}[htp]
    \centering
    \includegraphics[width=0.9\linewidth]{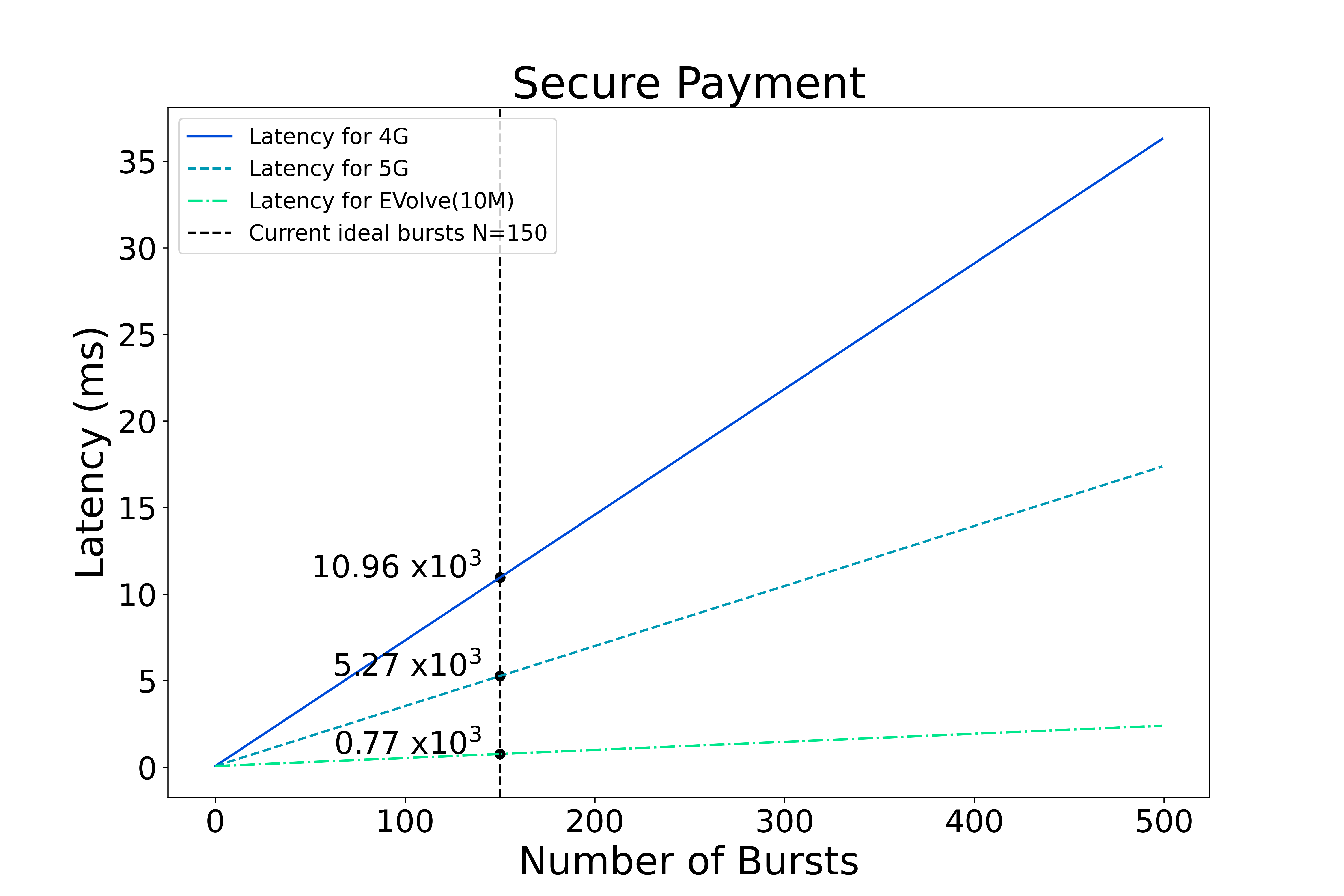}
    \caption{Latency for Secure Payment with an Increasing Number of Bursts.}
    \label{fig: Micropayments latency for an increasing number of bursts}
\end{figure}

The latency results for an increasing number of bursts are presented in Fig. \ref{fig: Micropayments latency for an increasing number of bursts}. It is important to note that all the PLC solutions overlap in curves and values. Thus, we chose to plot only the 10 Mbps option for clarity.

This scenario demonstrates the significant performance advantages of the Secure Payment VAS when utilizing a low-latency method like PLC. While both 4G and 5G exhibit a steeper slope in the latency curve—indicating greater variability and performance degradation over time—all PLC methods show a much flatter, lower-angled curve. This consistency underscores PLC’s ability to handle frequent, small transactions efficiently with minimal fluctuation, making it particularly suited for high-frequency communication scenarios. As a result, PLC enhances performance and provides a stable and secure foundation for applications like Secure Payment, where reliability and low latency are essential.

\subsection{Resource Overhead Analysis}

The overhead evaluation for \acronym demonstrated efficient resource usage and scalability. The EVerest charging stack \cite{everest_website} and EdgeX edge server \cite{edgexfoundry} contributed approximately 2 GB and 1.3 GB to the disk footprint, respectively, staying well within the 256 GB SD Card capacity. Dynamic monitoring using the `vmstat` utility confirmed stable RAM usage, with minimal swapping and over 95\% CPU idle time during steady-state operations. This indicates that \acronym operates reliably under its current workload and can accommodate additional VASs without significant performance degradation.

\subsection{Requirements Discussion}

The design of \acronym emphasizes the trade-offs between operation, performance, and security through the following considerations: First, the platform adheres to standard interfacing and communication protocols to support different suppliers across the ecosystem. This is achieved by leveraging ISO 15118 for standardized communication between the EV and EVSE, alongside MQTTS and HTTPS, implemented through EdgeX Foundry's Southbound and Northbound interfaces, ensuring seamless and scalable data exchange. Second, the architecture is designed to be layered, enabling the addition of new hardware components, such as GPUs, FPGAs, DPUs, and TEEs, as well as software modules and communication means. This extensibility is supported by EdgeX Foundry’s modular microservices, EVerest’s charging stack, Docker containerization for flexible deployment, and Redis for scalable caching, ensuring that application-layer development remains unaffected by infrastructure changes. Third, the platform incorporates best communication and computing security practices by employing TLS and OCPP in compliance with ISO 15118 standards, Vault for secure cryptographic key management, and OAuth for multi-factor authentication. Access Control Lists and MQTTS ensure privacy and controlled access, which safeguard data integrity and minimize exposure risks. Finally, resilience is supported by Consul, enabling fault-tolerant capabilities to ensure continuous operation even under adverse conditions. Together, these technologies form the foundation of our \acronym implementation, enabling a secure, scalable, and adaptable framework for the EV ecosystem.

\section{Conclusion}
\label{sec:conc}

The \acronym platform demonstrates the potential to transform EV chargers into secure, efficient, and scalable hubs for VASs by leveraging edge computing and advanced communication technologies. Evaluation results show that the emerging 100 Mbps PLC standard outperforms 4G and 5G in stability and throughput, making it ideal for latency-sensitive tasks like vehicle-to-charger communication. Secure payment VAS particularly benefits from PLC’s low latency and high bandwidth, providing a secure and efficient alternative to traditional methods. Additionally, vehicular software updates and SIEM applications exhibit significant improvements over 4G, even in a 10 Mbps PLC environment, demonstrating the platform’s scalability and adaptability. These results establish \acronym as a robust solution for advancing smart EV charging systems, addressing gaps in ISO 15118, and enabling secure, scalable VAS deployment while balancing edge and cloud resources for optimal performance.




\bibliographystyle{IEEEtran}
\bibliography{ref}

\end{document}